\documentclass[letter]{aa}  
\usepackage{graphicx}
\usepackage{txfonts}
\usepackage{hyperref}
\newenvironment{simplechar}{%
   \catcode`\$=12
   \catcode`\&=12
   \catcode`\#=12
   \catcode`\^=12
   \catcode`\_=12
   \catcode`\~=12
   \catcode`\%=12
}{}

\begin{document} 

   \title{Host galaxy properties of quasi-periodically erupting X-ray sources}

   \author{T. Wevers
          \inst{1}
          \and
          D.R. Pasham
          \inst{2}
          \and 
            P. Jalan 
          \inst{3}
          \and
          S. Rakshit
          \inst{3}    
          \and 
          R. Arcodia
          \inst{4}
          }

   \institute{European Southern Observatory, Alonso de C\'ordova 3107, Casilla 19, Santiago, Chile\\ \email{twevers@eso.org}
              \and
            MIT Kavli Institute for Astrophysics and Space Research, Cambridge, MA 02139, USA
            \and
            Aryabhatta Research Institute of Observational Sciences, Manora Peak, Nainital 263002, India
            \and
            Max-Planck-Institut für Extraterrestrische Physik (MPE), Giessenbachstrasse 1, 85748 Garching bei München, Germany
             }

   \date{}

% \abstract{}{}{}{}{} 
% 5 {} token are mandatory
 
  \abstract
  % context heading (optional)
  % {} leave it empty if necessary  
   {Quasi-periodic X-ray eruptions (QPEs) are a recently discovered phenomenon, the nature of which remains unclear. Based on their discovery in active galactic nuclei (AGN), explanations related to an AGN accretion disk, or potentially stellar tidal disruption event (TDE), were put forward. Following the report of QPEs in apparently passive galaxies, alternatives including highly unequal mass compact object binaries have been proposed to explain their properties.}
  % aims heading (mandatory)
   {We perform a systematic study of the five known QPE host galaxies with the aim of providing new insights into their nature.}
  % methods heading (mandatory)
   {We analyse new and archival medium resolution optical spectroscopy of the QPE hosts. We measure emission (and absorption) line fluxes, their ratios and equivalent widths (EWs), to locate the QPE hosts on diagnostic diagrams. We also measure the velocity dispersion of the stellar absorption lines to estimate their black hole masses.}
  % results heading (mandatory)
   {All QPE host galaxies show emission lines in their optical spectra. Based on their ratios and EWs, we find evidence for the presence of an active galactic nucleus in all sources, including those previously reported as passive. We measure velocity dispersions between 36 and 90 km s$^{-1}$, implying the presence of low mass (10$^{5 - 6.7}$ M$_{\odot}$) black holes, consistent with literature findings. Finally, we find a significant over-representation (2/5 sources, or a factor of 13$^{+13}_{-10.5}$) of quiescent, Balmer strong (post starburst) galaxies among QPE hosts.}
  % conclusions heading (optional), leave it empty if necessary 
   {The presence of a narrow line region consistent with an AGN in all QPE host galaxies implies that a pre-existing accretion flow likely plays an integral part to the QPE phenomenon. The strong over-representation of quiescent Balmer strong galaxies among QPE hosts can be naturally explained in both the TDE and interacting extreme mass ratio inspiral hypotheses.}
   \keywords{galaxies: active – galaxies: nuclei – quasars: general – quasars: supermassive black hole}

   \maketitle
%
%-------------------------------------------------------------------

\section{Introduction}
Quasi-periodic eruptions (QPEs) are recurring X-ray outbursts that have recently been discovered from some extragalactic nuclei (see \citealt{Miniutti19} for the first example). These large amplitude (in excess of an order of magnitude), short (duration of several 10s of minutes to hours) eruptions recur on timescales of hours to $\sim$ a day \citep{Miniutti19, Giustini20}. First discovered serendipitously in archival {\it XMM-Newton} observations, more recently they were also discovered in a blind X-ray survey performed by the eROSITA instrument \citep{Predehl2021} aboard the Spectrum-Roentgen-Gamma telescope \citep{Arcodia19} as well as targeted searches in X-ray archives \citep{Chakraborty21}.
 
This new class of events contains four known members and one candidate member\footnote{One source is considered a candidate QPE because only a small number of cycles have been detected so far.} to date.
Two sources (GSN069 and RXJ1301) were discovered in active galactic nuclei (AGN). Their amplitude and duration have an energy dependence, being stronger and shorter at higher energies. During quiescence, their X-ray spectrum is consistent with a featureless, cool (kT$\sim$40--60 eV) accretion disk model (e.g. \citealt{Miniutti19}), and the inferred black hole masses are $\sim$10$^{5-6}$ M$_{\odot}$. During the outbursts a second thermal component (with higher temperature, $\sim$120 eV) becomes apparent. In light of these properties, the first potential explanations that were put forward focused on accretion disk instabilities \citep{Miniutti19, Giustini20}, either related to a pre-existing AGN disk or to a long-lived (but recently formed) disk following the tidal disruption of a star. \citet{Sukova21} alternatively explore how an extreme mass ratio inspiral (EMRI) embedded in a gaseous accretion disk affects the accretion rate, finding results which are broadly consistent with QPE observed properties.

Subsequently, two QPEs were discovered in a blind X-ray survey (eRO-QPE1 and eRO-QPE2, \citealt{Arcodia19}). Finally, the newest candidate member of the QPE family is 2MASXJ0249, discovered in the host galaxy of a tidal disruption event (TDE) candidate \citep{Chakraborty21}.
The discovery of QPEs hosted in {\it apparently} passive galaxies \citep{Arcodia19} raises some obvious issues for the (pre-existing) AGN accretion disk scenarios, although in the TDE scenario this is not problematic. Furthermore, some properties of recently discovered QPEs (e.g. asymmetries between the fast rise and slower decay phases) cannot be well explained by accretion disk instabilities, at least according to our current understanding. 
Alternative hypotheses, which do not require the presence of an AGN accretion disk, have therefore also been proposed. These include extreme mass ratio inspirals (EMRIs) of (for example) a white dwarf (WD) orbiting a supermassive black hole (SMBH, \citealt{King20}) as well as more generally extreme mass ratio compact object binaries \citep{Arcodia19}, interacting EMRIs (i.e. multiple objects around a single SMBH, \citealt{Metzger21}), and self-lensing of massive SMBH binaries (although there are some inconsistencies between amplitudes and timescales in this model, \citealt{Ingram21}). 

Given that their nature remains unclear, in this Letter we report on the analysis of new and archival optical spectra of the QPE host galaxies. The aim of this work is to systematically characterise their host properties, and ultimately provide new insights into their nature. 

\section{Observations and analysis}
\label{sec:observations}
For three QPEs (GSN069, eRO-QPE1 and eRO-QPE2) we present new, medium resolution optical spectroscopy obtained with the Magellan Echelette spectrograph (MagE, \citealt{mage}), mounted on the Magellan Baade telescope at Las Campanas Observatory, Chile. The spectra were taken with a 0.7$^{\prime\prime}$ slit, which leads to a (slit-limited) spectral resolution of R$\approx$5900, equivalent to an instrumental FWHM broadening of 50 km s$^{-1}$ at 4000$\AA$. 
These observations are reduced using the dedicated MagE reduction pipeline \citep{Kelson00, Kelson03}.
The reduced spectra are subsequently normalised to the continuum by fitting a low order spline function for each order (where emission and absorption lines are excluded through masking and an iterative sigma-clipping routine) and stitched together (see Figure \ref{fig:host_magellan}, black, for an example); inverse-variance mean weighting is used for wavelength regions which are covered by multiple échelle orders. 

\begin{figure*}
    \centering
    \includegraphics[width=\textwidth]{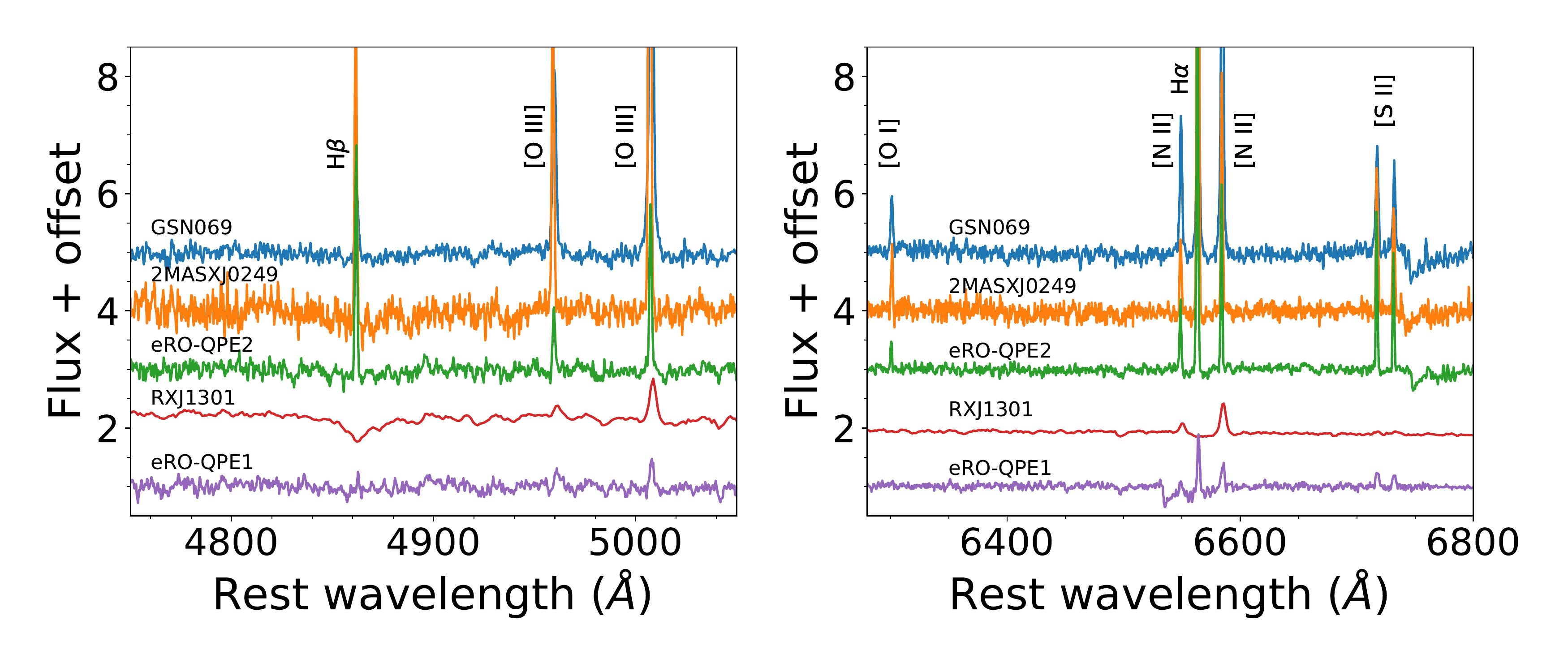}
    \caption{Insets of the host galaxy spectra of all 5 QPE sources. All spectra show emission lines typical of actively accreting and/or star-forming galaxies, although some emission lines are only visible after subtracting the stellar (galaxy) component.}
    \label{fig:spectra}
\end{figure*}

We also reanalyse an archival Keck/Echelette Spectrograph and Imager (ESI, \citealt{esi}) spectrum (for 2MASXJ0249, presented in \citealt{Wevers19a}) and an archival Sloan Digital Sky Survey (SDSS) spectrum (for RXJ1301, first presented in \citealt{Dewangan2000}). Details of all the spectra used in this work can be found in Table \ref{tab:observations}. Insets of the spectra, focusing on the emission lines used in the analysis, are shown in Figure \ref{fig:spectra}. 

\begin{table*}[]
    \caption{Details of the spectroscopic observations used in this work and main results. Redshifts are adopted from the literature. For the SDSS spectrum we report the radius of the fiber, rather than the slit width. The black hole mass is reported using the M--$\sigma$ relation of \citet{Gultekin09}. EW(H$\alpha$) lists the measured EW of the line in the rest frame, where values in brackets denote uncertainties on the last digit. For eRO-QPE1 and RXJ1301 values in square brackets denote measurements on the stellar absorption corrected (i.e. template subtracted) spectra; for RXJ1301 there is no emission before correcting stellar absorption. H$\delta_A$ represents the Lick H$\delta_A$ absorption index. Negative values indicate lines in emission.}
    \centering
    \begin{tabular}{c|cccccccc}
    Source & z & Telescope/instrument & Slit width & $\sigma_{\rm instr}$ & $\sigma_{\rm meas}$ & M$_{\rm BH, \sigma}$  & EW(H$\alpha$) & H$\delta_A$\\
     & & & (arcsec) & (km s$^{-1}$) & (km s$^{-1}$) & log$_{10}$(M$_{\odot}$) & ($\AA$) & ($\AA$)\\\hline
    
       GSN069  & 0.018 & Magellan/MagE & 0.7 & 21 &63$\pm$4 & 5.99$\pm$0.50 & --12.5(4) & 3.5$\pm$0.2\\
        RXJ1301 & 0.0237& SDSS & 3 & 70 & 90$\pm$2 & 6.65$\pm$0.42 & --- / [--1.7(3)] & 6.26$\pm$0.05\\
        eRO-QPE1 & 0.0505& Magellan/MagE & 0.7 & 21 & 56$\pm$5 & 5.78$\pm$0.55 & --2.8(4) / [--3.4(5)] & 2.7$\pm$0.3 \\
        eRO-QPE2 & 0.0175& Magellan/MagE & 0.7 &  21 & 36$\pm$3 & 4.96$\pm$0.54 & --13.1(2) & 1.1$\pm$0.4\\
        2MASXJ0249 & 0.019 & Keck/ESI & 0.5 & 16 & 43$\pm$4 & 5.29$\pm$0.55 & --33.2(5) & --2.9$\pm$0.9\\\hline
    \end{tabular}

    \label{tab:observations}
\end{table*}

\subsection{Velocity dispersion measurements}
To measure the host velocity dispersion of the stellar absorption lines, we use the pPXF software \citep{Cappellari2017} combined with the ELODIE \citep{Prugniel2001,Prugniel2007} spectral library (which has a spectral dispersion of 0.54$\AA$, slightly better than the MagE and ESI observations) to perform full spectrum fitting (masking strong emission lines originating from ionised gas) through convolution with a variable broadening kernel. For the (lower resolution) SDSS spectrum of RXJ1301, we use the standard MILES single stellar population templates \citep{Falcon2011}. 
The instrumental broadening is subtracted (in quadrature) before measuring the observed line broadening (represented by the velocity dispersion, $\sigma$) using all available absorption lines in the spectral range 3800--6800$\AA$. We resample the spectrum 1000 times within the uncertainties and perform a full spectrum fit on each realisation of the data. The reported velocity dispersion and uncertainty are the mean and standard deviation of the resulting distribution, which closely follows a Gaussian distribution. We show an example of the best-fit solution for eRO-QPE1 in Figure \ref{fig:host_magellan}, and the resulting measurements can be found in Table \ref{tab:observations}.

\subsection{Emission and absorption line measurements}
To measure (narrow) emission line fluxes, we use the {\tt lmfit} Python package \citep{lmfit} to create a simple model including a polynomial (to account for the continuum flux) and a total of 9 Gaussian components, one for each prominent narrow emission line. No broad lines are apparent in the spectra, and no complicated deblending is required. Due to the difficulty in obtaining accurate absolute flux calibrations for echelle spectrographs, and since we are interested primarily in their equivalent widths (EWs) and line ratios, we measure the emission line fluxes and EWs on the continuum-normalised spectra. The continuum normalisation is straight-forward in all cases, and we estimate (by artificially increasing/decreasing the continuum level) that this introduces systematic uncertainties of $<25\%$ in the measurements. 

The weakness of H$\beta$ relative to H$\alpha$ in eRO-QPE1 and RXJ1301 (for the latter, see \citealt{Dewangan2000}) indicates that this line is likely affected by stellar absorption. This is typical for post-starburst galaxies, which contain a significant population of A stars (leading to strong stellar absorption in the Balmer lines). For these two sources, we therefore subtract the best-fit stellar template obtained with pPXF to account for this effect (see Figures \ref{fig:sub_qpe1} through \ref{fig:sub_rxj1301_zoom} for visual illustrations), before measuring the (narrow) emission line fluxes. 
Lick H$\delta_A$ absorption indices, optimised for stellar absorption from A stars, are measured as defined in \citet{Worthey1997}. The main results are presented in Table \ref{tab:observations}.

\section{Results}
\label{sec:results}
\subsection{The host galaxies of QPEs (very) likely host an active nucleus}
\begin{figure*}
    \centering
    \includegraphics[width=\textwidth]{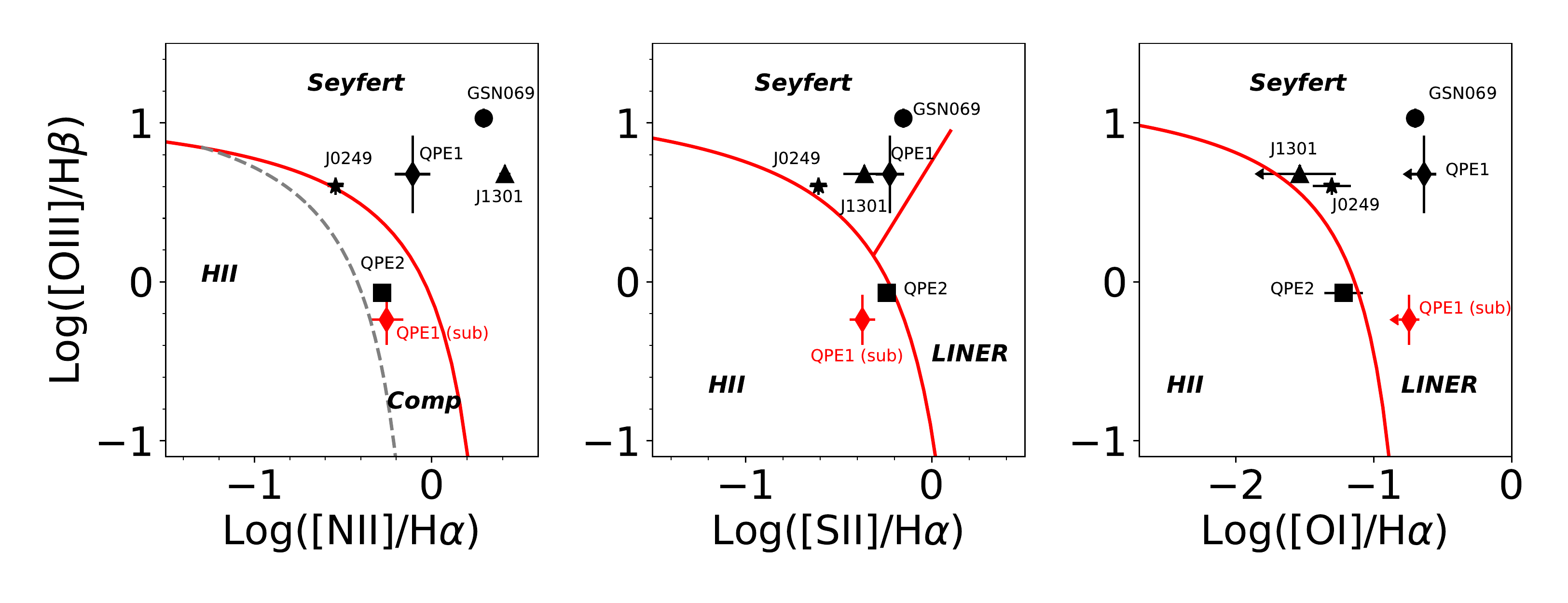}
    \caption{BPT diagram of the five QPE host galaxies. All sources are beyond the pure star-formation (black dashed) line, indicating that they (very) likely host an (in some cases low luminosity) active galactic nucleus. The data to reproduce this plot is provided in Table \ref{tab:ratios}.}
    \label{fig:bpt}
\end{figure*}

\begin{figure}
    \centering
    \includegraphics[width=\columnwidth]{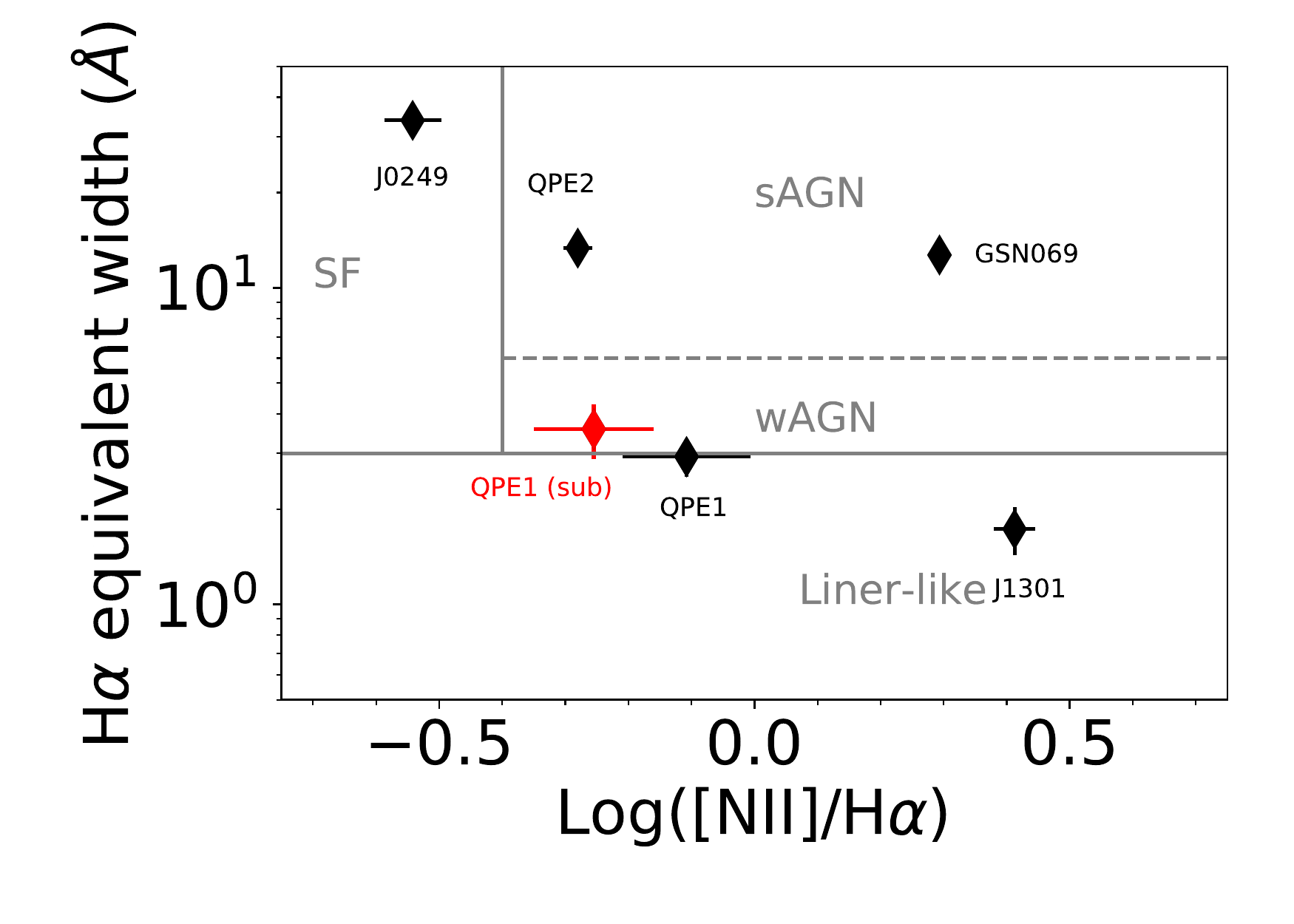}
    \caption{Diagnostic diagram of the five QPE host galaxies based on H$\alpha$ EW and relative [N\,\textsc{ii}] / H$\alpha$ strength, based on the selection criteria presented in \citet{Cid2011}. RXJ1301 (a post-starburst galaxy) is located among the {\it LINER-like} population (due to the relative weakness of the H$\alpha$ EW), whereas in the BPT diagram it falls firmly in the AGN locus. 2MASXJ0249 is classified as dominated by star formation, whereas it falls in the Seyfert part of the BPT diagram. The data to reproduce this plot is provided in Table \ref{tab:ratios}.}
    \label{fig:whan}
\end{figure}

The host galaxies of each of the five known QPEs show prominent narrow emission lines, although some of these only become apparent after subtracting a template for the stellar component of the galaxy. Using the measured (narrow) line fluxes, we create diagnostic diagrams to assess the dominant source of ionising photons powering the emission lines. We present the Baldwin-Phillips-Terlevich (BPT) diagram \citep{bpt} in Figure \ref{fig:bpt}, and overlay the classical lines that separate different classes of sources according to the dominant ionisation mechanism \citep{Kauffmann2003, Kewley2001}: purely star-forming, composite (star-forming + AGN) and AGN (Seyfert/LINER) galaxies. Three out of five sources (2MASXJ0249, GSN069 and RXJ1301) lie within the Seyfert part of the diagrams. eRO-QPE2 is well beyond the extreme starburst line and not affected by any stellar absorption, so it is highly likely that this source contains an AGN (see also Figure \ref{fig:whan}). eRO-QPE1, finally, is a more complex case, given the strong stellar absorption. The black diamond shows the measurements without correcting for this effect, while the red diamond shows the results from the template subtracted spectrum. In both cases the source is outside of the pure star-forming region, indicating that it probably also contains an AGN. 

In Figure \ref{fig:whan} we show the EW of H$\alpha$ versus [N\,\textsc{ii}/H$\alpha$] (WHAN) diagnostic diagram \citep{Cid2011}. Here we note that three out of five sources, including eRO-QPE1 and eRO-QPE2, fall within the AGN region. RXJ1301 (a strong post-starburst galaxy) is located among the retired/low ionisation nuclear emission line region (LINER)-like galaxies, which is common for galaxies with relatively weak emission lines \citep{Cid2011}. This may be a consequence of its edge-on orientation \citep{Caldwell1999}. There is ample evidence that this source indeed hosts an AGN (e.g. \citealt{Dewangan2000, Giustini20}). 2MASXJ0249, finally, is within the star-formation dominated region of this diagram.

Our medium resolution optical spectroscopy of in particular eRO-QPE1 differs significantly from the spectrum presented in \citet{Arcodia19}. While no emission lines were clearly visible in their (low resolution) spectrum, we clearly detect emission lines of H$\beta$, H$\alpha$ as well as the typical galaxy emission lines of [O\,\textsc{iii}], [N\,\textsc{ii}] and [S\,\textsc{ii}]. We speculate that this is due to a combination of higher spectral resolution and smaller slit aperture of our Magellan spectra; the latter leads to less contamination due to non-nuclear galaxy light. The same effects may also be responsible for the different emission line ratios measured for eRO-QPE2, which we find falls within the composite region in the BPT diagram and the AGN region of the WHAN diagram, rather than in the star-forming region as reported by \citet{Arcodia19}. Remeasuring the line fluxes in the case of 2MASXJ0249 also leads to slightly different results compared to \citet{Wevers19a}, who reported line measurements performed in {\tt iraf} which likely have larger systematic errors than the method used in this work.

We conclude from Figures \ref{fig:bpt} and \ref{fig:whan} that our optical spectroscopy provides strong evidence for an AGN in 4/5 sources. We remark that there remain some (well-known) discrepancies between the BPT and WHAN diagram classification for 2MASXJ0249. In this respect, the WHAN diagram provides the {\it optimal} dividing line between SF and AGN dominated line ratios, and as such is not a strict division (i.e. it is well known that galaxies hosting an AGN can fall to the left this line, and vice-versa for star formation dominated galaxies). For the final source (eRO-QPE1) there are more systematic uncertainties, but we deem it probable that it also contains an AGN, as the lines cannot be explained by star formation alone. We also note that the narrow line region diagnostics are sensitive to changes in black hole activity only on timescales of several 100--10$^5$ years (their size in light travel time), and we do not detect broad emission line components in any source. 

\subsection{The host galaxies of QPEs have low mass black holes}
The velocity dispersion measurements can be used to estimate the black hole mass by exploiting the well-known M$_{\rm BH}$ -- $\sigma$ correlation. The measured velocity dispersions (Table \ref{tab:observations}) correspond to black hole mass estimates of 10$^{5-6}$ M$_{\odot}$ for 4/5 sources, the exception being RXJ1301 for which we infer log$_{10}$(M$_{}$) = 6.5$\pm$0.4 (0.3) M$_{\odot}$ using the the M-$\sigma$ relation from \citet{Gultekin09}, or log$_{10}$(M$_{}$) = 7.0$\pm$0.3 M$_{\odot}$ with the \citet{Kormendy13} relation. Regardless of the specific calibration of the M--$\sigma$ relation that is used (and the associated uncertainties in their calibration at such low values for $\sigma$), such low velocity dispersions exclude the presence of massive ($>$10$^7$ M$_\odot$) black holes, which are typical for the AGN population at large (e.g. \citealt{Woo2002, Rakshit2020}), for the majority of QPE hosts. 

\subsection{Quiescent Balmer strong galaxies are over-represented in QPE hosts}
\begin{figure}
    \centering
    \includegraphics[width=\columnwidth]{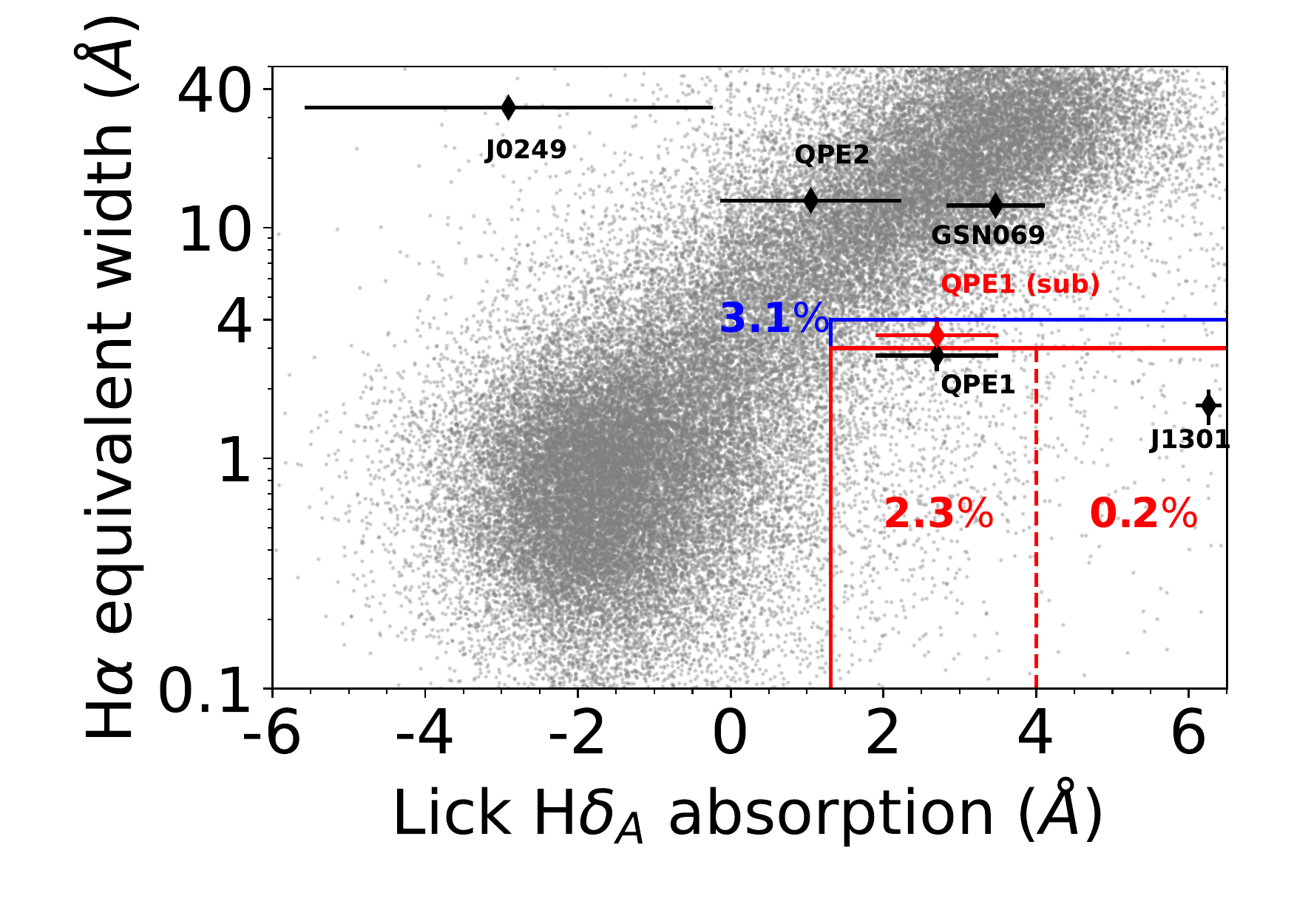}
    \caption{Spectral properties of SDSS galaxies (grey dots) and QPE hosts (diamonds). 2/5 QPE hosts lie in a sparsely populated region of this diagram of H$\alpha$ EW vs. Lick H$\delta_A$ absorption index. The relative fraction of sources within the coloured boxes is shown: E+A galaxies represent only 0.2$\%$ of SDSS galaxies (black dashed box), while quiescent Balmer strong galaxies represent 2.3$\%$ (black solid box). The slightly widened (blue) box that encompasses eRO-QPE1 (H$\alpha$ EW $<$ 4 $\AA$) contains 3.1$\%$ of SDSS galaxies. }
    \label{fig:lick}
\end{figure}

In Figure \ref{fig:lick} we plot the QPE hosts among the SDSS galaxy population in terms of H$\alpha$ EW (in the rest-frame) and Lick H$\delta_A$ absorption index\footnote{See Appendix \ref{sec:sql} for the SQL queries used to define this sample and the relative fractions reported later.} from the MPA-JHU catalog \citep{Brinchmann2004}. Comparing the location of the QPE hosts to the SDSS galaxy population shows that two sources fall within the {\it blue cloud galaxies} with both moderate H$\alpha$ emission and H$\delta_A$ absorption. Furthermore, it stands out that 2/5 QPE hosts (eRO-QPE1 and RXJ1301) occupy a very sparsely populated region of the diagram at low H$\alpha$ EW and strong H$\delta_A$ absorption. This region is home to post-starburst (E+A) galaxies (typically defined as having H$\alpha$ EW $<$3$\AA$ and Lick absorption index H$\delta_A$ - $\delta_{H\delta_A} > 4 \AA$) and relatedly, quiescent Balmer strong (QBS) galaxies (typically defined as H$\alpha$ EW $<$3$\AA$ and Lick absorption index $>$1.31$\AA$, although we note that the H$\alpha$ EW requirement is somewhat arbitrary). The H$\alpha$ EW criterion selects galaxies with little on-going star formation (i.e. well below the star formation main sequence), while the absorption index criterion selects galaxies with a significant population of A-stars, thought to have formed in a recent ($<$1 Gyr) starburst. To quantify the rarity of the parameter space where eRO-QPE1 and RXJ1301 are found, we select SDSS galaxies (from DR17) with H$\alpha$ EW $<$4$\AA$ and Lick absorption index $>$1.31$\AA$ (i.e. slightly more generous than the typical QBS selection) and find that these represent 3.1 $\%$ of the total population\footnote{For reference, E+A galaxies make up $\sim 0.2 \%$ of the total population and QBS galaxies with H$\alpha$ EW $<$3$\AA$ make up 2.3$\%$, see e.g. \citet{French17}.}. The presence of 2/5 = 40$\%$ of the total QPE sample within this region therefore indicates a significant over-representation, by a factor of 13$^{+13}_{-10.5}$ (i.e. $> 2.5$, adopting binomial small number 95$\%$ confidence limits, \citealt{Gehrels1986}) of these galaxies among the QPE hosts compared to SDSS galaxies. The false alarm probability of randomly drawing 2 sources inside this sparsely populated parameter space is $\sim 0.1 \%$.

\section{Discussion and conclusions}
\label{sec:discussion}
Black hole masses derived from X-ray spectral fitting, as well as the observed X-ray luminosities and estimates based on galaxy scaling relations indicate that the black hole masses of QPEs are in the range 10$^5$ -- few $\times$10$^6$ M$_{\odot}$. Our velocity dispersion measurements further suggest that, regardless of the uncertainties in the absolute calibration of the M--$\sigma$ scaling relations at such low measured values, the black hole masses of QPE hosts are $<$ 10$^7$ M$_{\odot}$, and in some cases hint at M$_{\rm BH}$ as low as 10$^5$ M$_{\odot}$. While this is atypical for the local galaxy population, it is very similar to the host galaxies of TDEs \citep{Wevers17, Wevers19a}. The significant over-representation of rare QBS galaxies among QPE hosts provides a further intriguing link to TDE host galaxies \citep{Arcavi2014, French2016}. In addition, one of the QPE candidates (2MASXJ0249) is indeed a TDE candidate (\citealt{Esquej2007, Auchettl2017}, although not one with a QBS host). In summary, 4/5 QPEs have a (indirect) connection (through their long-term X-ray properties or their host galaxy) with the TDE population. Identifying and studying a larger sample of QPEs will reveal whether this is the consequence of small number statistics, selection biases, or whether a true connection between QPEs and TDEs can be established.

One clear difference with the TDE host galaxy population are the emission line ratios, as most TDEs (both UV/optical and X-ray selected) are discovered in passive galaxies without emission lines or in star forming galaxies (e.g. \citealt{French2020, Sazonov21}). Our measurements suggest that the QPE host galaxies contain an ionising source in addition to star formation.
The implication is that all QPE host galaxies likely host an active galactic nucleus. If the association between TDEs and QPEs (based on their host galaxy properties) is real, this may suggest that QPEs are related to special/exotic TDE scenarios, e.g. as the result of a partial disruption and the subsequent capture of the stellar remnant, which interacts with a pre-existing AGN accretion flow (e.g. \citealt{Xian21}). 
Differences in selection methods between X-ray and UV/optical surveys of TDEs could also contribute to this effect; for example, TDE follow-up observations are more likely to target quiescent galaxies than AGNs.

We note that the QPE quiescent X-ray spectra can be well described by a featureless accretion disk model \citep{Arcodia19, Chakraborty21}, which is consistent with the presence of an active nucleus. However, as pointed out in detail by \citet{Arcodia19}, current models that invoke radiation pressure instabilities to explain QPE properties do not explain some aspects satisfactorily, including the need for artificially high viscosities or black hole masses which are very strongly incompatible (factor $\sim$1000) with estimates based on galaxy scaling relations presented in this work and the literature.

Alternative hypotheses to explain QPEs include extreme mass-ratio compact object binaries \citep{King20, Arcodia19}. \citet{King20} argue that a compact object (white dwarf) on a highly eccentric orbit, following a partial stellar disruption and subsequent in-orbit capture of the surviving remnant (i.e. the QPE phenomenon only lasts for $\sim$ several 1000 years), can potentially explain the properties of GSN069. 
However, \citet{Metzger21} find that the parameter space for single EMRIs is so small that it is hard to reconcile with the observed QPE occurrence rate, and argue instead that the interaction between two EMRIs is a more likely scenario.

In the single EMRI scenario, it is not clear why both targeted searches in X-ray archives (e.g. \citealt{Chakraborty21}), as well as blind searches would discover QPEs exclusively around AGNs, which make up only a small minority of galaxies ($\sim$ 5--10 per cent, e.g. \citealt{Lopes2017}) in the local Universe. The caveat here is that a sample size of 5 sources is too small to make conclusive claims at present. We do note that AGN activity is associated with post starburst evolution, which peaks with a delay of $\sim$250 Myrs after the starburst \citep{Wild2010}; \citet{Pawlik18} estimate an AGN duty cycle of $\sim$50$\%$ during the post starburst phase. This could indicate that the QPE phase preferentially occurs shortly after the starburst is triggered. The presence of residual emission lines in the 2 post-starburst QPE hosts could indicate that the bursts occurred recently and are still being quenched. Determining the detailed post-starburst ages of a (larger) sample of QPE hosts could help to elucidate this potential preferential timing and connection.

\citet{Metzger21} argue that the active phase of an interacting EMRI system is long enough to potentially create its own narrow line region, and furthermore that the presence of a gaseous AGN disk may favour the migration of circular EMRIs into the galactic nucleus. Hence this latter scenario is consistent with our finding that a narrow line region is present in all five known QPEs, whereas the single EMRI scenario is not. \citet{Metzger21} also argue that post-starburst (or at least, post merger) galaxies might be favoured QPE hosts due to the presence of migration traps as a result of AGN gas, which can produce a sequence of co-orbiting EMRIs. 
Finally, the discovery of multiple periodicities in eRO-QPE1 (Pasham et al., in preparation) could be naturally explained in the multiple EMRI scenario. 

More generally, it remains ambiguous at present what is the exact link between the narrow emission line diagnostics and the intrinsic ionising source. The lack of broad emission lines could indicate that none of the SMBHs is currently actively accreting, i.e. all 5 QPEs have been discovered in systems with recently shut-down AGN. A different scenario (given the archival X-ray detections of at least 2 sources, GSN069 and RXJ1301) is that these galaxies host low luminosity AGN where no luminous broad line region is present, and/or our observations are not sensitive enough to detect the weak broad line region. Finally, if QPEs are a long-lived phenomenon capable of injecting a significant amount of energy into the SMBH surroundings, they could power their own NLR without the need for a {\it traditional} AGN.\\ 

We end by briefly summarising the main results of this work. We have presented a systematic analysis of optical spectroscopic observations of the five known QPE host galaxies. This analysis provides new insights regarding the nature of this phenomenon:
\begin{enumerate}
    \item We have confirmed the low mass nature of the QPE host black holes through velocity dispersion measurements, consistent with the X-ray properties of the QPEs.
    \item We have established the presence of (narrow) emission lines in all QPE host galaxies. Based on their line ratios and equivalent widths, a narrow line region consistent with being (at least partially) ionised by an AGN is present in all QPEs. It remains unclear whether this narrow line region can be powered by QPEs alone (as postulated in some scenarios), or requires a long-lived AGN.
    \item QPE explanations invoking accreting processes (AGNs/TDEs, or a combination of both), as well as interacting EMRI scenarios are compatible with the reported host galaxy properties.
    \item We have presented a strong over-representation of (rare) quiescent Balmer strong galaxies among the QPE hosts, by a factor of 13$^{+13}_{-10.5}$. This provides an intriguing, even if currently unclear, link to the host galaxies of tidal disruption events. The interacting EMRI scenario also provides a natural explanation for this over-representation.
\end{enumerate}

We end by noting that the line ratios presented in this study are also measured from ground-based optical spectroscopy. These measurements are known to be affected by finite aperture effects and galaxy light contamination due to limitations imposed by atmospheric seeing. Nevertheless, studying the host galaxy properties of a larger sample of QPEs may help to further elucidate their nature in the future.

\begin{acknowledgements}
This paper includes data gathered with the 6.5 meter Magellan Telescopes located at Las Campanas Observatory, Chile (P.I. Pasham). The spectra will be made publicly available upon publication.
\end{acknowledgements}
\bibliography{aanda}{}
\bibliographystyle{aa}

\appendix
\onecolumn
\section{Supplementary figures and tables}
\begin{figure}[h!]
    \centering
    \includegraphics[width=\textwidth]{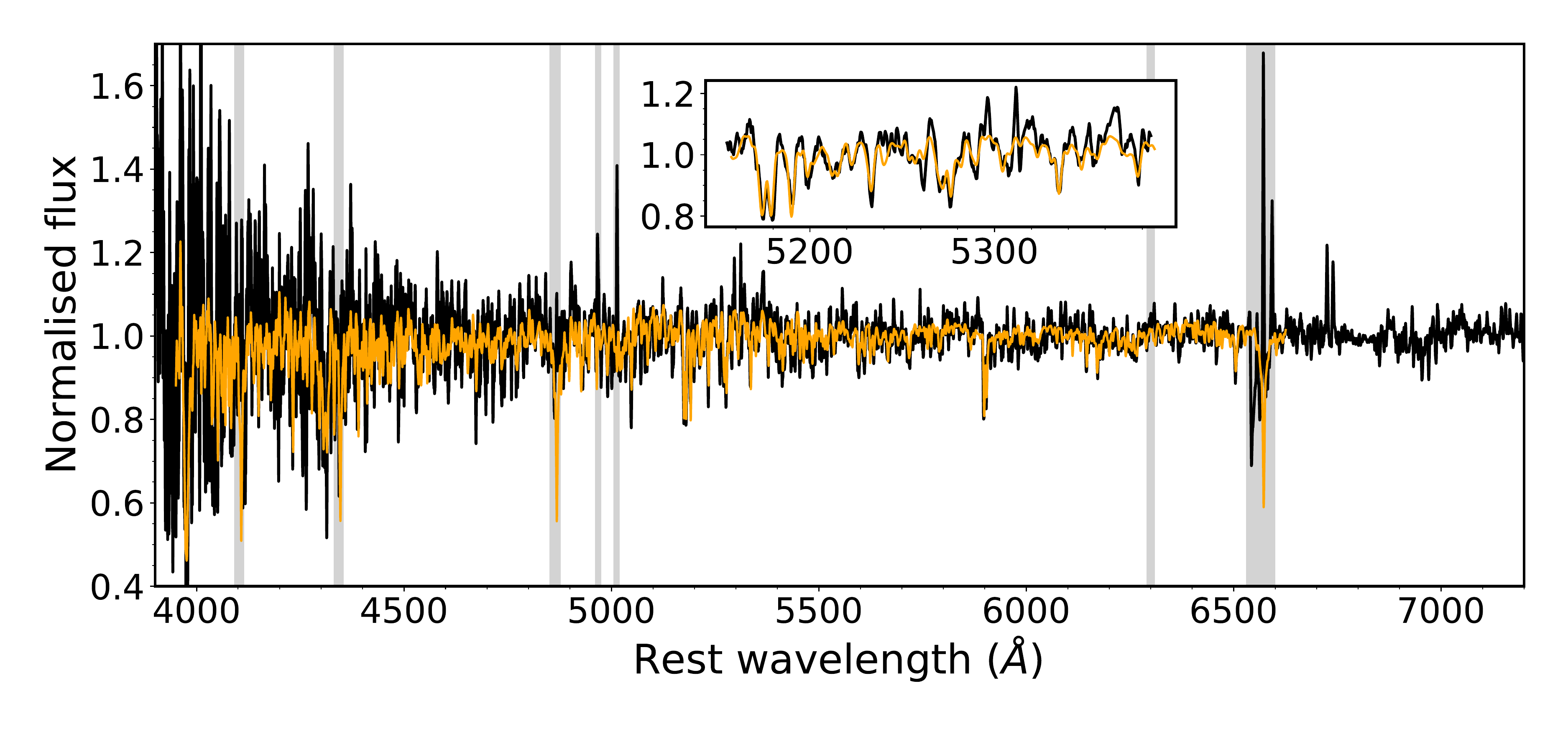}
    \caption{Magellan/MagE spectrum (black) of the host galaxy of eRO-QPE1. Overlaid in orange is the best-fit template that is used to measure the host velocity dispersion. The inset shows the region around the Mg\,\textsc{i} b triplet to illustrate the quality of the fit. Gray bands indicate host galaxy emission lines, which are excluded from the fit. The spectrum is convolved with a Gaussian kernel of width 5 pixels for clarity.}
    \label{fig:host_magellan}
\end{figure}

\begin{figure}
    \centering
    \includegraphics[width=\textwidth]{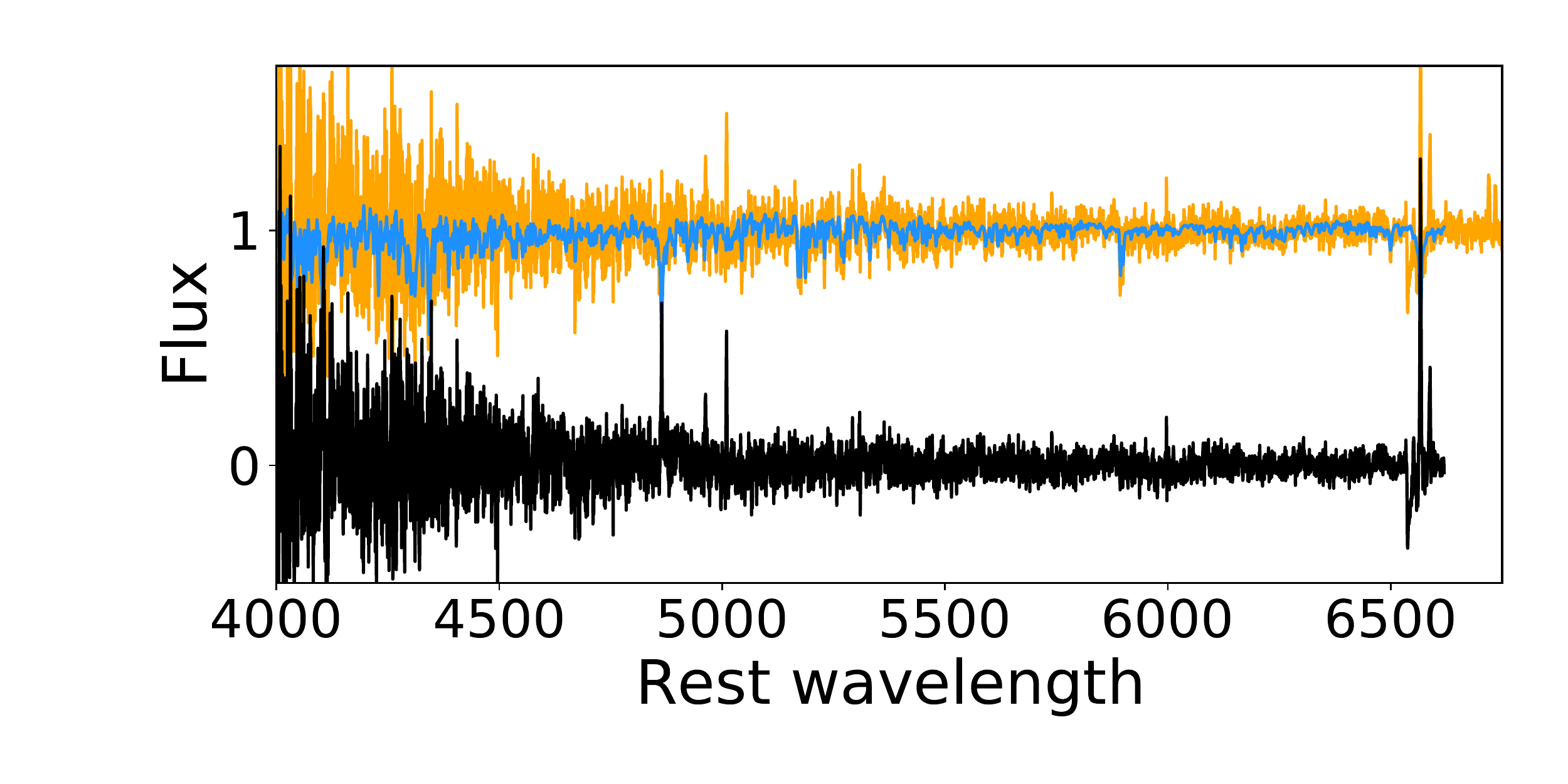}
    \caption{Best-fit template (blue) overlaid on the data (orange) and the result of the template subtraction (black) for the host galaxy of eRO-QPE1. }
    \label{fig:sub_qpe1}
\end{figure}

\begin{figure}
    \centering
    \includegraphics[width=\textwidth]{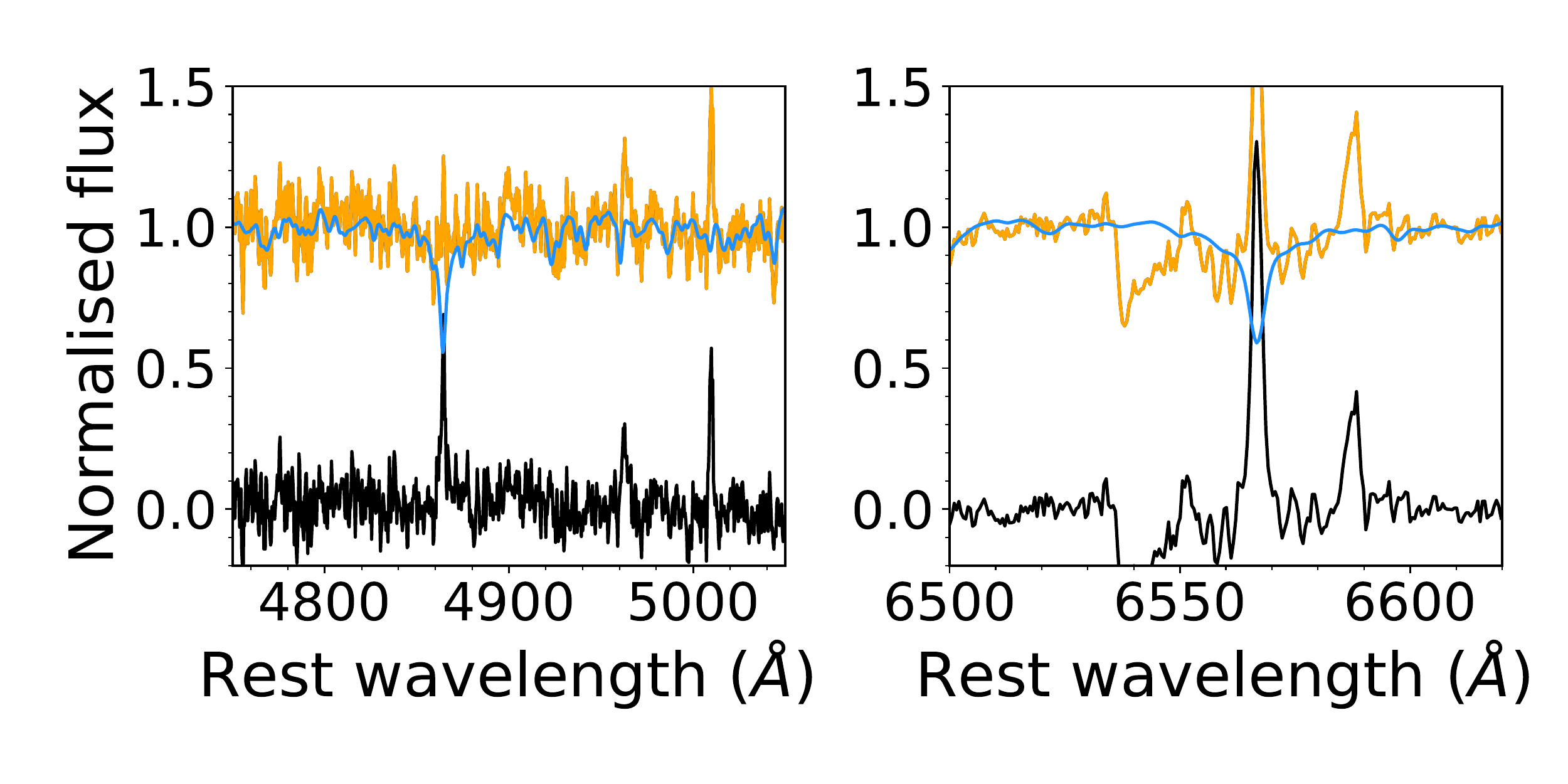}
    \caption{Zoom-ins of the H$\beta$ (left) and H$\alpha$ (right) regions of the best-fit template (blue) overlaid on the data (orange) and the result of the template subtraction (black) for the host galaxy of eRO-QPE1. }
    \label{fig:sub_qpe1_zoom}
\end{figure}

\begin{figure}
    \centering
    \includegraphics[width=\textwidth]{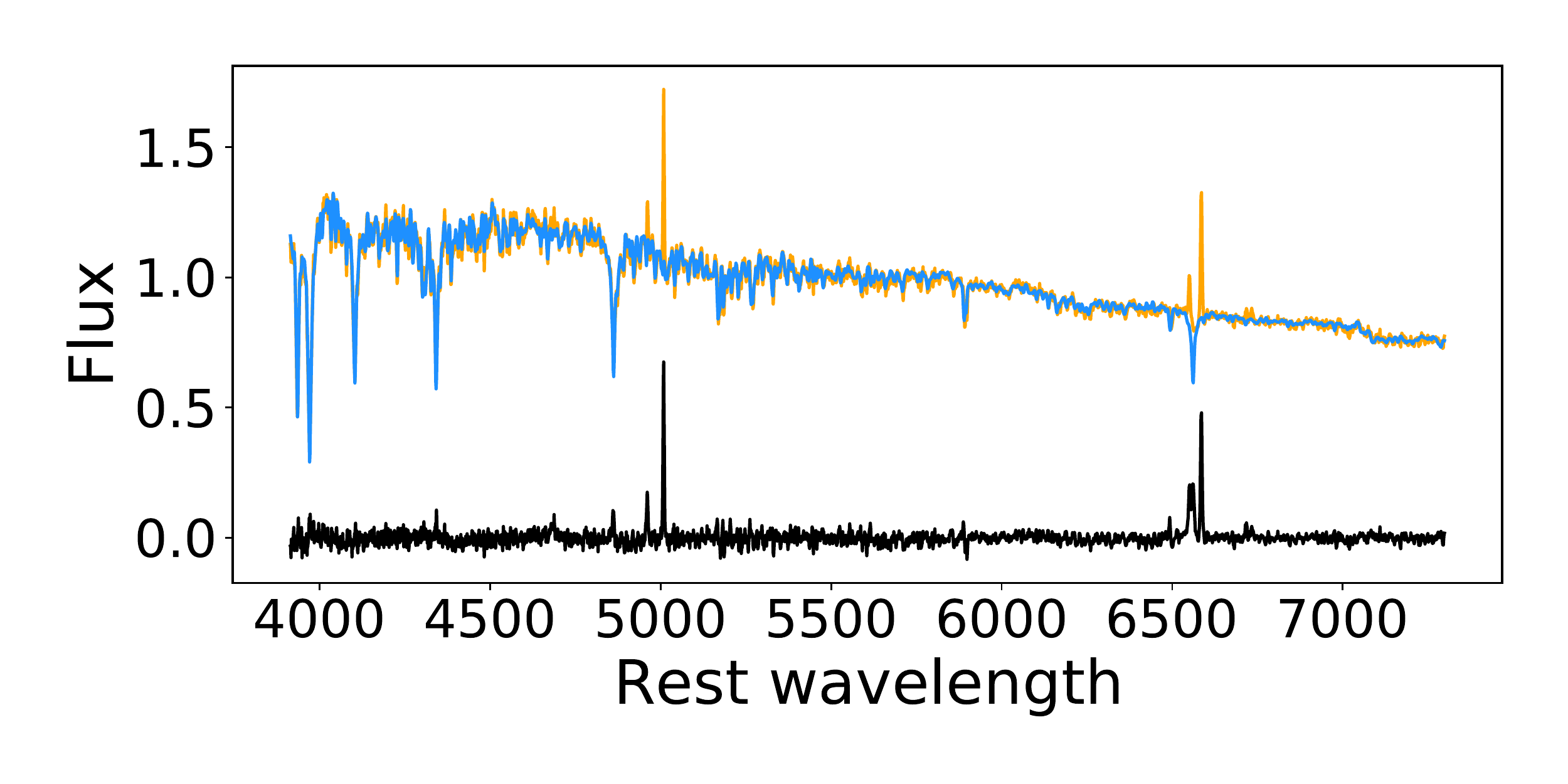}
    \caption{Best-fit template (blue) overlaid on the data (orange) and the result of the template subtraction (black) for the host galaxy of RXJ1301. }
    \label{fig:sub_rxj1301}
\end{figure}

\begin{figure}
    \centering
    \includegraphics[width=\textwidth]{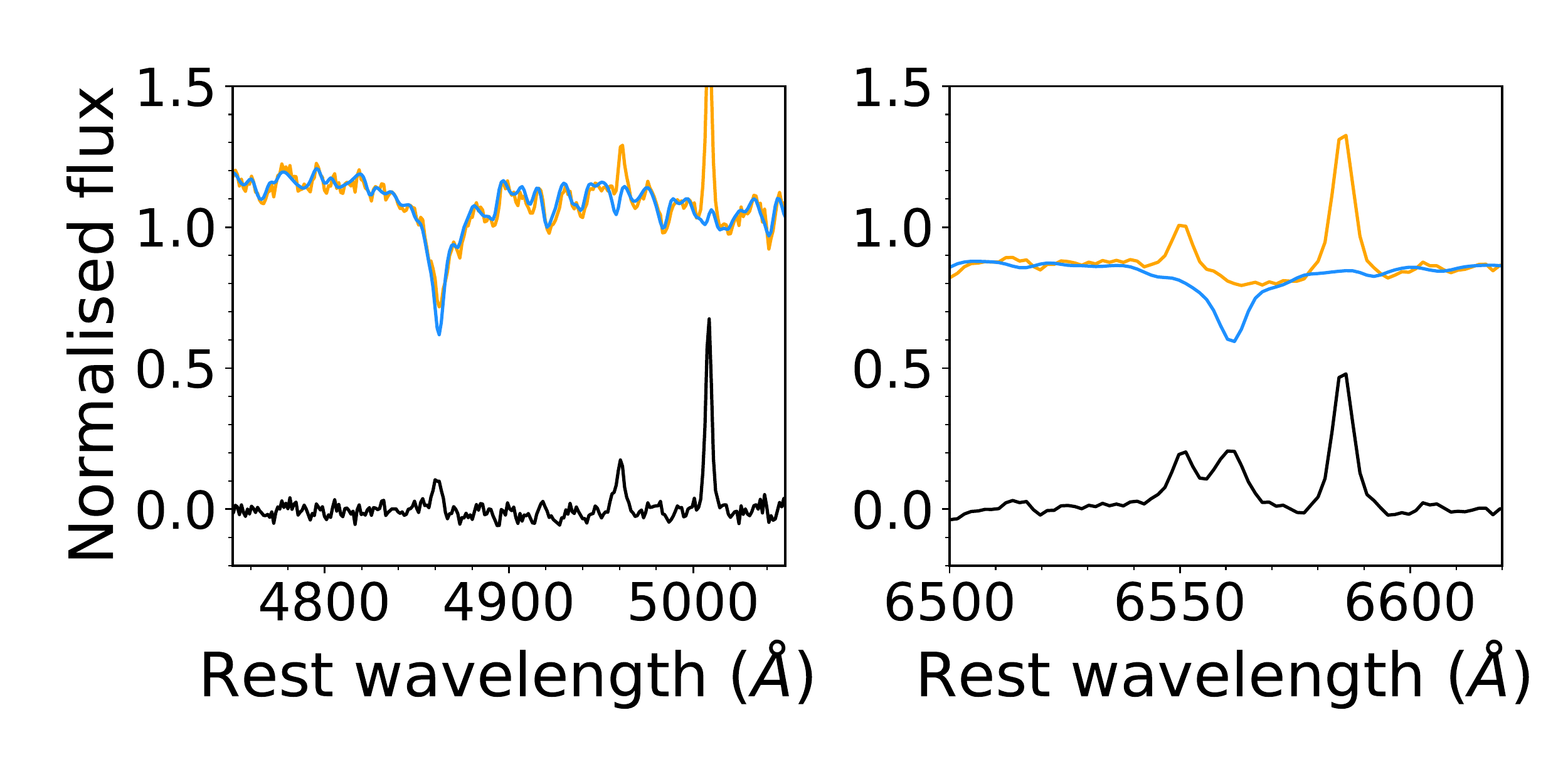}
    \caption{Zoom-ins of the H$\beta$ (left) and H$\alpha$ (right) regions of the best-fit template (blue) overlaid on the data (orange) and the result of the template subtraction (black) for the host galaxy of RXJ1301. }
    \label{fig:sub_rxj1301_zoom}
\end{figure}

\begin{table*}[]
    \caption{Line ratios measured from the spectra. For eRO-QPE1 and RXJ1301 values in square brackets denote measurements on the stellar absorption corrected (i.e. template subtracted) spectra; for RXJ1301 there is no emission before correcting stellar absorption.}
    \centering
    \begin{tabular}{c|cccc}
    Source & log$_{10}$(N\,\textsc{ii} / H$\alpha$) & log$_{10}$(S\,\textsc{ii} / H$\alpha$) & log$_{10}$(O\,\textsc{i} / H$\alpha$) &log$_{10}$(O\,\textsc{iii} / H$\beta$) \\\hline
    
    GSN069  & 0.29 $\pm$0.02 & --0.15$\pm$0.03 & --0.70$\pm$0.05 & 1.03$\pm$0.06\\
    RXJ1301 & [0.41$\pm$0.03] & [--0.36$\pm$0.11] & [$<$--1.5] & [0.68$\pm$0.05]\\
    eRO-QPE1 & --0.11$\pm$0.1 [--0.25$\pm$0.1] & --0.23$\pm$0.08 [--0.37$\pm$0.07] & $<$--0.63 [$<$--0.75] & 0.68$\pm$0.24 [--0.24$\pm$0.16]\\
    eRO-QPE2 & --0.28$\pm$0.02 & --0.24$\pm$0.02 & --1.22$\pm$0.0.14 & --0.07$\pm$0.02\\
    2MASXJ0249 & --0.54$\pm$0.05 & --0.60$\pm$0.05 & $<$--1.3 & 0.60$\pm$0.05\\\hline
    \end{tabular}

    \label{tab:ratios}
\end{table*}
\clearpage

\section{SQL queries}
\label{sec:sql}
\subsection{SQL query for a clean parent sample}
The following SQL query selects galaxies with a median signal-to-noise ratio of at least 10 to discard noisy line measurements. Only galaxies at redshift $>0.01$ are selected to remove galaxies which are much larger than the SDSS fiber. Unreliable H$\alpha$ EW width measurements are excluded by requiring 
$\rm h\_alpha\_eqw\_err > -1$. This yields a sample of 580182 galaxies.\\

\begin{simplechar}
SELECT\\
  COUNT(*), s.lick_hd_a, s.lick_hd_a_err,\\
    g.h_alpha_eqw, g.h_alpha_eqw_err\\
FROM    GalSpecLine AS g\\
  JOIN galSpecIndx AS s ON s.specobjid = g.specobjid\\
  JOIN SpecObjAll AS k ON k.specobjid = g.specobjid\\
WHERE\\
    h_alpha_eqw_err > -1\\
    AND k.class = 'GALAXY'\\
  AND k.snMedian >= 10\\
  AND K.Z > 0.01
\end{simplechar}

\subsection{SQL query for the E+A sample}
This SQL query incorporates the selection criteria for E+A post-starburst galaxies. It yields a sample of 1204 galaxies (0.2$\%$ of the parent sample).\\

\begin{simplechar}
SELECT\\
  COUNT(*)\\
FROM    GalSpecLine AS g\\
  JOIN galSpecIndx AS s ON s.specobjid = g.specobjid\\
  JOIN SpecObjAll AS k ON k.specobjid = g.specobjid\\
WHERE\\
  h_alpha_eqw_err > -1\\
  AND h_alpha_eqw > -3\\
  AND lick_hd_a - lick_hd_a_err > 4\\
  AND k.class = 'GALAXY'\\
  AND k.snMedian >= 10\\
  AND K.Z > 0.01\\
\end{simplechar}

\subsection{SQL query for the QBS sample}
This SQL query incorporates the selection criteria for QBS galaxies (with $\rm h\_alpha\_eqw > -3$) and returns a sample of 14407 galaxies (2.3$\%$ of the parent population). Relaxing the criterion to $\rm h\_alpha\_eqw > -4$ to conservatively incorporate QPE1 in this sample results in a sample of 18023 galaxies (3.1$\%$ of the parent sample).\\

\begin{simplechar}
SELECT\\
  COUNT(*)\\
FROM    GalSpecLine AS g\\
  JOIN galSpecIndx AS s ON s.specobjid = g.specobjid\\
  JOIN SpecObjAll AS k ON k.specobjid = g.specobjid\\
WHERE\\
  h_alpha_eqw_err > -1\\
  AND h_alpha_eqw > -3\\
  AND lick_hd_a > 1.31\\
  AND k.class = 'GALAXY'\\
  AND k.snMedian >= 10\\
  AND K.Z > 0.01\\
\end{simplechar}

\end{document}